\documentclass[sigconf, table, dvipsnames]{acmart}
\usepackage{listings}
\usepackage{booktabs}
\usepackage{array}
\usepackage{multirow}
\usepackage{enumitem}
\usepackage[linesnumbered,ruled,vlined]{algorithm2e}
\usepackage{lineno}
\DeclareUnicodeCharacter{2264}{\ensuremath{\leq}}

\renewcommand\footnotetextcopyrightpermission[1]{} 

\AtBeginDocument{%
  \providecommand\BibTeX{{%
    \normalfont B\kern-0.5em{\scshape i\kern-0.25em b}\kern-0.8em\TeX}}}

\usepackage[acronym]{glossaries}
\makeglossaries
\newacronym{pjia}{PJIA}{Princess Juliana International Airport}
\newacronym{OC}{OC}{Offloading Coordinator}
\newacronym{GSE}{GSE}{Ground Support Equipment}
\newacronym{oc}{OC}{Offloading Coordinator}
\newacronym{gse}{GSE}{Ground Support Equipment}
\newacronym{movcon}{MOVCON}{Movement Control}
\newacronym{atc}{ATC}{Air Traffic Control}
\newacronym{uld}{ULD}{Unit Load Device}
\newacronym{icu}{ICU}{Intensive Care Unit}
\newacronym{pax}{PAX}{Passengers}

\graphicspath{{./images/}} 
\begin{document}

\title{Effects of Unplanned Incoming Flights on Airport Relief Processes after a Major Natural Disaster}

\author{
  Luka Van de Sype$^{1}$, 
  Matthieu Vert$^{2}$*, 
  Alexei Sharpanskykh$^{2}$,
  and Seyed Sahand Mohammadi Ziabari$^{3}$
}

\affiliation{%
  \institution{
  $^{1}$Deerns, Anna van Buerenplein 21F 2595 DA, Den Haag, The Netherlands}
  \country{}
}

\affiliation{%
  \institution{
  $^{2}$Aerospace Engineering Faculty, Delft University of Technology, Kluyverweg 1, 2629 HS, Delft, The Netherlands}
  \country{}
}

\affiliation{%
  \institution{$^{3}$Department of Computer Science and Technology, SUNY Empire State University, Saratoga Springs, NY, USA}
   \country{}
}

\email{ lukavandesype@gmail.com, m.p.j.vert@tudelft.nl, o.a.sharpanskykh@tudelft.nl, sahand.ziabari@sunyempire.edu}

\begin{abstract}

{
The severity of natural disasters is increasing every year, having an impact on many people's lives. During the response phase of disasters, airports are important hubs where relief aid arrives while people need to be evacuated out of there. However, the airport often forms a bottleneck in these relief operations, because of the sudden need for increased capacity. 
Limited research is done on the operational side of airport disaster management. Experts identify the main problems as first the asymmetry of information between the airport and the incoming flights, and second the lack of resources.
The goal of this research is to gain understanding of the effects of incomplete knowledge of incoming flights with different resource allocation strategies, on the performance of the cargo handling operations in an airport after a natural disaster event.
An agent-based model is created, where realistic offloading strategies with different degrees of information uncertainty are implemented. Model's calibration and verification are performed with experts in the field.
The model performance is measured by the average turn-around time, which can be split in offloading time, boarding time and the cumulative waiting times. The results show that the effects of one unplanned aircraft are negligible. However, all waiting times increase with the more arriving unplanned aircraft.}

\end{abstract}



\fancyhead{}
\maketitle
\section{Introduction}
\label{ch:Intro}

Many natural disasters happen every year. Between 1994 and 2013 it was recorded that  a total of 6,873 natural disasters occurred worldwide \cite{human_cost_natural_disasters}. At the same time, due to climate change, severity of these natural disasters is increasing every year \cite{climate_disaster_1}. Many of those natural disasters already have a negative impact on many people's lives. Per year around 218 million people are affected by these disasters and on average 68,000 die \cite{human_cost_natural_disasters}. Airports are important hubs during the response phase of a humanitarian crisis. However, during the response phase often a bottleneck arises in the ground handling operations of the airport as there is a sudden increase in incoming cargo while not all resources of the airport are available because of the disaster \cite{airportcongestion_2018,Feil2018}. At the same time, many people need to be evacuated out of the area, adding complexity to the operations. Lives can be saved if these ground handling operations are executed more efficiently. \\

Even though lives are at stake, according to Polater \cite{Polater_2018}, airport disaster management is still in its infancy in terms of academic research. The authors show that especially in the operational and business side of airport disaster management there is an academic gap. Disaster management is often a retrospective analysis of all the events. While research on airport disaster management is mainly focused on disasters that happen on the airport and not the airport as a hub for a larger disaster event. A close research domain is disaster resilience. Disaster resilience is 'the capacity of a system, community or society potentially exposed to hazards to adapt, by resisting or changing in order to reach and maintain an acceptable level of functioning and structure' \cite{unisdr-resilience}. 

For the airport system, both disaster management and disaster resilience focus on strategies and interventions to build efficient emergency operations. However, there is a lack of studies that formally and systematically investigate the effects of resource allocation strategies on the aircraft handling operations performance, in particular in a context of limited information regarding incoming aircraft bringing support to the island's inhabitants.
One of the few exceptions is the van Liere's study \cite{GaelThesis}, that simulates the cargo handling and passenger evacuation operations during the response phase on the Saint Martin Princess Juliana International Airport (\gls{pjia}) after the hurricanes in 2017 using the agent-based modelling paradigm. 

Agent-based modelling and simulation (ABMS) allows researchers to formally represent real-world complex systems using interacting agents, and investigate the systems' behaviours through scenarios that are computationally simulated. Agents are autonomous entities with complex believes (e.g.: goals, plans) that can make decisions based on their perceived information and act on their environment. This micro-level modeling of agent behaviors aligns with approaches in \cite{passengerABM}, where individual passenger activities within terminals are shown to collectively impact overall system performance.
ABMS also connect system local properties to system global properties, such that the actions of agents can be causally connected to the overall performance of the system. In particular, agents' coordination \cite{Fines2020} and anticipation \cite{Blok2018} of potential or effective disturbances can be modelled and investigated with ABMS. Airports are complex sociotechnical systems and the extensive use of ABMS in the literature to model airport operations have shown that is a suitable approach  \cite{Bouarfa2013, Blom_2015,passengerABM, sanders2021agent,deBosscher2024, debosscher2023comprehensive}.

ABMS will be used in this study. Moreover, the model proposed in this work focuses on three problems: lack of information on incoming flights, lack of resources and lack of planning. 
As little research has been done on this topic, expert knowledge is essential for the model development. According to experts, in similar disaster relief there are generally three main reasons for delays: the arrival of unannounced flights, the airport not being equipped for handling large cargo aircraft and the preparation of airport management for such a disaster \cite{weeks}. These issues were also considered in van Liere's case study of Saint Martin island. \\

As the case study of Saint Martin addresses many general problems that occur during and after natural disasters, in particular related to lack of resources and information, this case will also be used in this research. van Liere's model simulates the cargo handling and passenger evacuation. This can be used to gain better understanding of the general bottlenecks that occur during airport cargo handling after a natural disaster event. This paper will focus on the offloading of civilian aircraft, as the airport focuses largely on civilian operations and the military, and civilian operations are often separated from each other. Furthermore, van Liere's study does not provide a detailed analysis of civilian operations.

The amount of knowledge possessed by airport employees of the incoming flights can create a situation of asymmetry of information. If the flight schedule is fully known, the airport operations coordinator can make decisions for resources allocations integrating aircraft time of arrivals, aircraft types, and cargo types. If the schedule is partially known or completely unknown the predictions made by the operation coordinator, based on the corresponding decisions, may not be efficient. 

The research objective is to gain understanding of the effects of incomplete knowledge of incoming flights, with different resource allocation strategies, on the performance of the cargo handling operations in an airport after a natural disaster event. 

Answering this question will provide more understanding of what offloading coordinator decisions are the best. In order to fulfil the research objective, first the agent-based model needs to be developed to realistically represent an airport after a natural disaster event. The main goals and performance metrics of the offloading coordinators need to be established, in order to know how the performance of cargo handling should be measured. Furthermore, it should be determined what resource allocation strategies are used for offloading civilian aircraft. Then, the influence of lack of information on incoming flights on the performance of the system can be analysed by performing agent-based simulations.

More specifically, the following methodological steps are taken: first to gain the necessary background information for the research and understand the case study, in addition to understanding the existing model and its limits. This is addressed in Section \ref{sec:CaseStudy}. 
The second step is the development of the agent-based model, based on the existing van Liere's model. The changes to the existing model and the agent-based model for this research are explained in Section \ref{sec:PJIA}. In Section \ref{sec:PJIA_prelim_analysis} the agent-based model is analysed and calibrated. The third step is to create and implement realistic scenarios with resource allocation strategies for offloading. These are more extensively explained in Section \ref{sec:methodology}. In Section \ref{sec:Results} the results of the experiments are given and discussed. Lastly, Section \ref{sec:Concl} gives the conclusion of this research.

\section{Background Information and Case Study}

\label{sec:CaseStudy}
The context of airport disasters, and in particular the case of Saint Martin in 2017, are detailed in this section.

\subsection{Airports and disasters}
\label{ss:CS_airportdistaster}
A humanitarian disaster is \textit{``an event or series of events that represents a critical threat to the health, safety, security, or well-being of a community or other large group of people, usually over a wider area."} \cite{UN-def}.
This paper focuses on sudden onset natural disasters, there are \textit{``triggered by a hazardous event that emerges quickly or unexpectedly."} \cite{DisTerminilogy}.
Examples of such a disaster is the Tsunami in Tonga in the beginning of 2022, or the hurricanes in Saint Martin in 2017.  \\

Disaster management is often described in a cycle of four phases: mitigation, preparedness, response, and recovery \cite{Cycle4phases}. This paper will focus on the response phase, which starts during or right after the disaster event. Reduction or elimination of the negative effects is the goal of this phase. It is in this phase when the airport forms a bottleneck as a extensive relief aid needs to go to the impacted areas, but the means are not always present \cite{coppola2007DMC,Berry2009-DMC}. 
Logistics is a large part of these humanitarian operations, and it has become a popular research field for the passed ten years. The \textit{Humanitarian Supply Chain} however, is mostly looked at from an overall point of view and not only the airport is considered \cite{information-book, short_relevantresearchHumDis}.

To gain understanding of the general problems related to the airport during the response phase, an interview was held with an expert on airport disaster logistics. Delays were identified as one of the essential problems, and the main causes the expert identified for delays can be summarised as \cite{weeks}:
\begin{enumerate}
    \item Scheduled flights are often not a problem, but the extra unscheduled flights cause a bottleneck, especially when the content of the cargo is unknown. 
    \item Island holiday destinations are mostly not equipped for handling large cargo aircraft, as normally most cargo comes in by ship or belly cargo.
    \item Not all countries and airports have plans in case there is a disaster and part of the airport is destroyed. Moreover if there is one it is often not executed as it is specified. 
\end{enumerate}

\subsection{Case Study: Saint Martin 2017}
\label{ss:CS_StMaarten}

On September $6^{th}$ of 2017 a category 5 hurricane on the Saffir-Simpson scale called Irma passed Saint Martin. This is the highest category on this scale. The hurricane reached wind speeds of 260 $km/h$ \cite{irmaspeed}. There were 4 fatalities and 23 injuries and over 70$\%$ of the buildings on the islands were destroyed. This affected a lot of people and the total damage cost was estimated to be around 1.5 billion US Dollars \cite{ReportIrma}.
On September $8^{th}$ and September  $18^{th}$, two other hurricanes passed the Caribbeans called Jose and Maria respectively, they did not go over Saint Martin but caused tropical storms \cite{ReportJose, ReportMaria}.

On the first day after hurricane Irma, all efforts were put in making the airport operational again.
There was a lot of destruction causing waiting times at the airport. The main issues were similar to the general problems identified by experts \cite{weeks}:
\begin{itemize}
    \item There was no electricity, mobile connection, internet and civil air traffic control. Hence, there was no knowledge of incoming civilian aircraft and all communication was done face to face.
    \item Most buildings were destroyed, as well as equipment and aircraft. Meaning there was a lack of resources, equipment and personnel.
    \item As the airport management did not show up, there was a lack of overall coordination.
\end{itemize}
As there was no electricity and in the beginning of such operations gas is scarce \cite{weeks, braam}. All communication was done face to face and everybody had to walk to exchange information and coordinate.

Furthermore, at that time in Saint Martin, the military could use only one of the two highloaders as the safety barrier was in the way when going under the aircraft. For one of the highloaders the safety barrier was taken off \cite{GaelThesis}.

As the case study fits the general identified problems that occur during and after disasters, it makes it a suitable case study to investigate.

\section{PJIA Agent-based Model}
\label{sec:PJIA}

In this section a model for Princess Juliana International Airport(PJIA), is described. The PJIA model is based on an existing model: the van Liere's model \cite{GaelThesis}. 

\subsection{General overview of the model}
\label{sss:CS_assump_lim}
In van Liere's model, three different operations were considered: civilian offloading, military offloading, and evacuation operations.

The last two operations will be simplified in the base model, as the focus lies on civilian offloading. 

The influence of military offloading on the civilian offloading operations is mainly through the use of equipment and the use of parking spots. Therefore, the military aircraft are still modeled, but their offloading is represented as a time penalty. 

Evacuation operations always occurs after offloading is finished. The impact of the evacuation on the civilian offloading operations is the time when the tasks after evacuation can start, as well as the time when the parking spot is used. For this reason the evacuation operations is only modelled as a time penalty.

According to the experts, the main goal of the OC is to lower the turn-around time (TAT). Subsequently, in these meetings the resource allocation strategies, used in at PJIA and in general, were established as the following \cite{weeks,braam,nancy,pilot}:
\begin{itemize}
    \item Aircraft are offloaded in chronological order of arrival.
    \item Maximum set of resources are deployed to offload an aircraft.
    \item If it is known that another aircraft is coming in, this will be taken into account in order to allocate the resources.
    \item Even though the general rule is offloading in chronological order, priority also plays a role. This can be a humanitarian priority such as an aircraft with medical equipment. It can also be business priority such as a regular client, as contractors are still a business. 
\end{itemize}
In the model on offloading strategy is implemented. The base model will have a maximal available resource strategy, as, according to the experts, this is a more appropriate representation of the case study. This means that for each available cargo door, one civilian handling agent with GSE is sent and that each civilian handling agent can have up to two drivers with tug and dollies.

\subsection{PJIA Model: Environment specifications}
\label{ss:PJIA_environment}
\begin{figure*}[h!]
\centering
\includegraphics[width=0.95\textwidth]{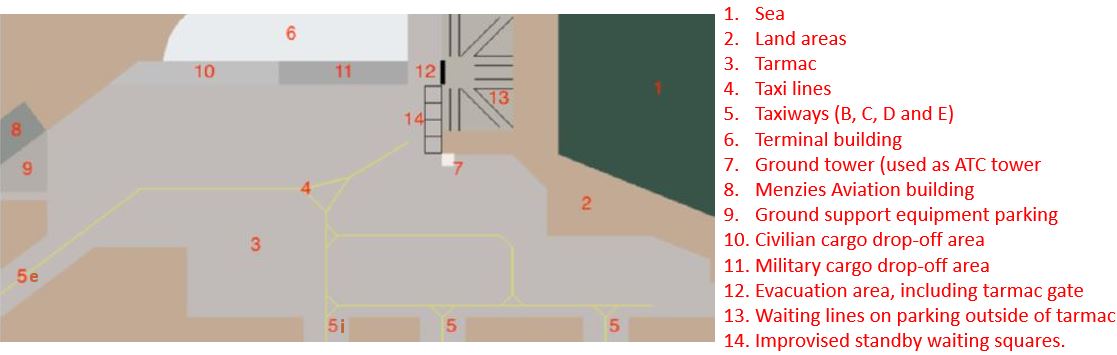}
\caption{Princess Juliana International Airport layout}
\label{fig:PJIA_lay_out}
\end{figure*}

The layout of the modelled airport environment can be found in Figure \ref{fig:PJIA_lay_out}. 

Aircraft come in from taxiway 5i. They taxi over the taxi lines to a parking spot. The military parking spots are in the top part of the tarmac, the civilian parking spots are at the bottom part. The aircraft exit through the most left taxiway, taxiway 5e. The \gls{OC}, civilian handling agents and drivers can be found in the Menzies Aviation building at 8. At 9 all the \gls{gse} can be found. The military offloading coordinator is standing on the tarmac in between 9 and 10. In front of the terminal building, 10 and 11 are cargo drop off areas for civilian and military cargo respectively. The \gls{movcon} can most of the time be found in 11. The military \gls{atc} agent can mostly be found at 7, the Ground Tower. 

\noindent One step time in the model is equal to 5 seconds. The environment also includes objects, there are cargo objects and \gls{GSE} objects. \\

\subsubsection{Cargo Objects} 
There are three types of cargo objects. After offloading, the cargo from civilian or military aircraft is put in the civilian or military drop off area respectively. 
\begin{itemize}
    \item One \textit{LD3 \gls{uld}} has an approximate weight of 1600 kg. They can be found in medium to large civilian aircraft.
    \item One \textit{463L Master Pallet} has an approximate weight of 4500 kg. They can be found in large military aircraft. 
    \item One \textit{\gls{icu}} has an approximate weight of 500 kg. This represents cargo in the form of boxes, packages and loose items. They can be found in all aircraft.\\ 
\end{itemize}

\begin{figure*}[h!]
\centering
\includegraphics[width=0.85\textwidth]{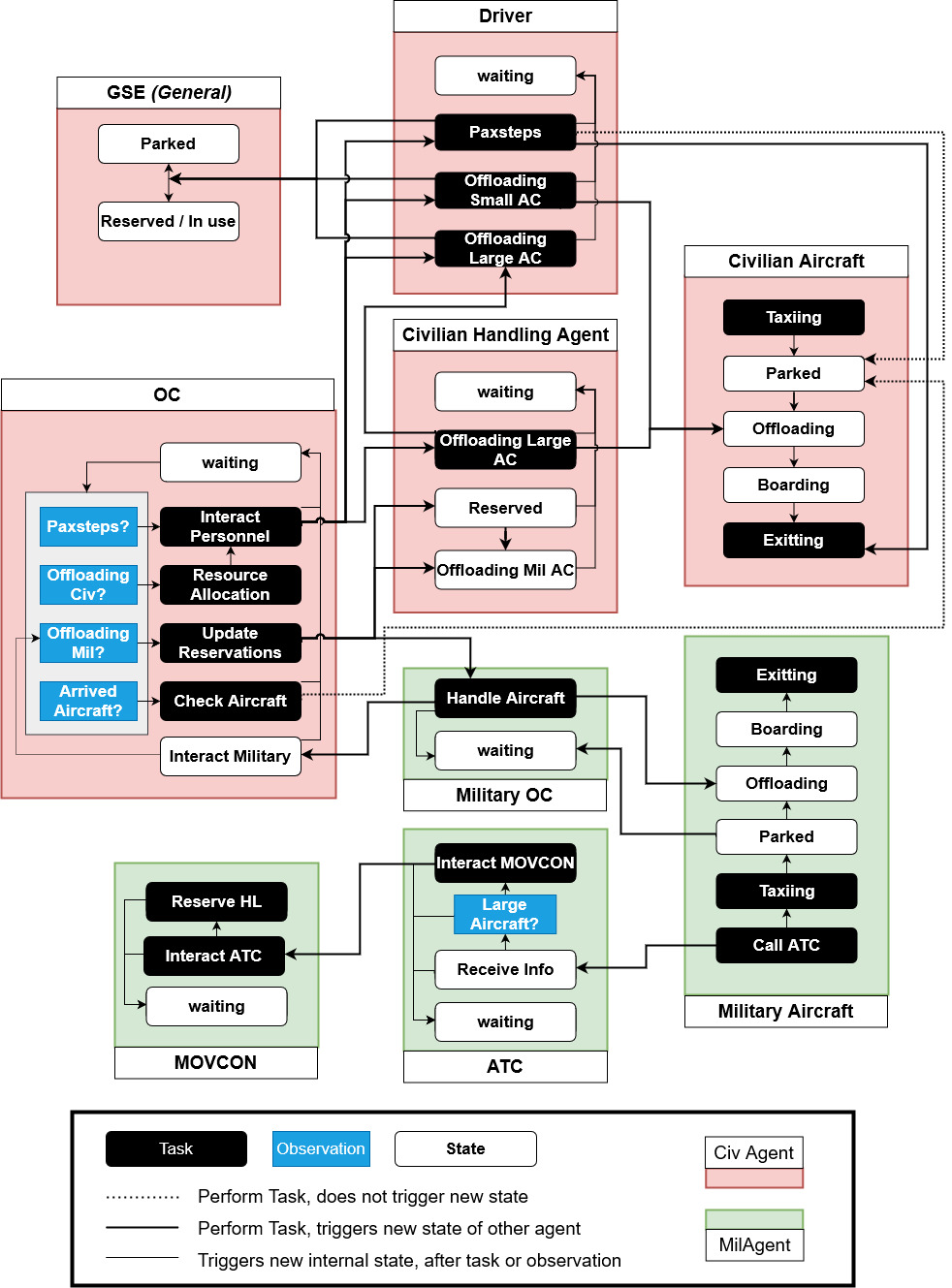}
\caption{Overview of the workings and interactions of the agents in the PJIA Model.}
\label{fig:PJIA_model_overview}
\end{figure*}

\subsubsection{Ground Support Equipment Objects}
All GSE vehicles move with the maximum allowed speed on a tarmac: 30 $km/h$. The GSE can be found in the ground support equipment parking. There are four types in the model:

\begin{itemize}
    \item \textit{Pax steps} are used on large civilian aircraft to get passenger on the aircraft, as well as for the aircraft crew to come down or the \gls{oc} to go up.
    \item \textit{Tugs with dollies} are the carts that bring the cargo from the aircraft to the terminal building.
    \item \textit{Belt loaders} are used to get ICU from large aircraft onto the dollies.
    \item \textit{Highloaders} are used to get LD3 ULD and 463L Master Pallet onto the dollies. However, only one of the two highloaders can be used on military aircraft. In the model the highloader which was adapted for military use is the one with the highest model ID number.
\end{itemize}
In general, a GSE has two states: it is either parked or in use. The change in state is caused by a driver or a civilian handling agents, using the GSE. In case the GSE is used for military purposes, the GSE also has a `reserved' state. This state is induced by the OC.

\subsection{PJIA Model: Agent specifications}
\label{ss:PJIA_agents}
In this section the properties, states and interdependencies of all agents are explained. Figure \ref{fig:PJIA_model_overview} gives an overview of the model.  As the equipment are objects, it was deemed not necessary to elaborate all of them in the figure. Therefore they are presented as general in the model overview. 

\subsubsection{Aircraft} \label{sss:aircraft_prop} 
Each aircraft crew and their aircraft was modelled as one agent. There are two types of aircraft agents: the Civilian Aircraft agent and the Military Aircraft agent. The arrival time of the agents is determined by the flight schedule, as explained in Section \ref{ss:meth_experiments}.


\paragraph{\textit{Call ATC Property}} This is only done by military aircraft. 15 minutes before arrival the aircraft calls the ATC and gives following information: its agent ID, arrival time, types of cargo, amount of cargo per type.

\paragraph{\textit{Taxiing Property}} The aircraft moves over the taxilines from its starting point to its destination point, using a predefined sequence of nodes.\\
ARRIVAL: Arrival always starts at taxiway 5i and its destination point is a free parking spot closest to the \gls{oc} building.\\
EXIT: Exit starts from the parking spot to the destination point at taxiway 5e.

As mentioned in Section \ref{ss:PJIA_environment}, the military and civilian parking spots are different. As overflow measure in the model, it is possible for civilian aircraft to park on a military parking spot in case no civilian parking spots left. Vice versa for the military aircraft.

\paragraph{\textit{Parked Property}} The \gls{oc} comes by and checks the aircraft content. The information the \gls{oc} needs is: type of aircraft (large/small), types of cargo and amount of each cargo type. The OC leaves, the aircraft's status is changed to `checked'. If the aircraft carry cargo, they wait for offloading.

\paragraph{\textit{Offloading Property}} If a large civilian aircraft carries cargo, then it will interact with the civilian handling agents. For a small civilian aircraft the interaction is directly with the driver. At each step, the civilian handling agent or driver is aware of the remaining cargo in the aircraft that they need to offload. They are only aware of another cargo type in the aircraft after they finish, as they can see if the aircraft is empty or not. If it is empty they communicate to the aircraft that offloading is finished.

For military aircraft a time penalty per kilograms of cargo is implemented. This timer starts at the moment when all the \gls{gse} and personnel are present at the aircraft. 

\paragraph{\textit{Boarding Property}} When \gls{pax} can be evacuated with the aircraft, a time penalty is implemented per pax.\\


\begin{figure}[h!]
\centering
\includegraphics[width=0.9\linewidth]{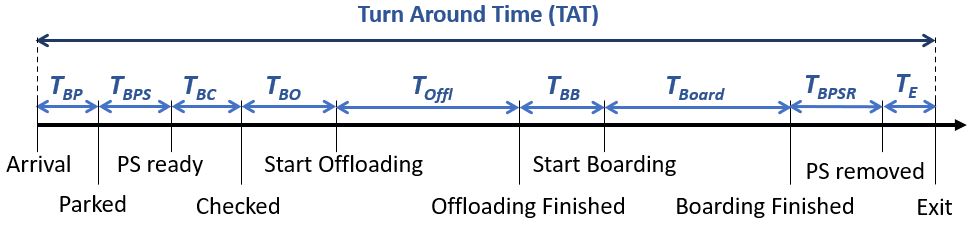}
\caption{Timeline of the turn-around process of a Civilian Aircraft.}
\label{fig:timeline_ac}
\end{figure}

Decreasing the turn-around time of civilian aircraft is the main goal of the OC, therefore an overview  of  the timeline of civilian aircraft is given in Figure \ref{fig:timeline_ac}. The TAT is defined from the moment the aircraft arrives until it exits the taxilane. In table \ref{tab:timeline_aircraft_explanation}, a brief explanation for all the time abbreviations is given. In this model the time between offloading and boarding is set to 12 minutes, which is equal to the average \textit{T$_{BB}$} from van Liere's model. The \textit{T$_{E}$} is negligible and will not be mentioned again. In this research these times are always mentioned in minutes.


\begin{table}[h!]
\caption{Explanation of all times in the aircraft timeline.}
\label{tab:timeline_aircraft_explanation}
\centering
\begin{tabular}{ll}
\multicolumn{1}{l|}{\textbf{Time}}    & \multicolumn{1}{l}{\textbf{Explanation}}     \\ \hline
\multicolumn{1}{l|}{$T_{BP}$}    & \multicolumn{1}{l}{Before Parked Time}     \\ 
\multicolumn{1}{l|}{$T_{BPS}$}  & \multicolumn{1}{l}{Before Paxsteps Ready Time}           \\ 
\multicolumn{1}{l|}{$T_{BC}$}   & \multicolumn{1}{l}{Before Cargo Content is Checked Time} \\ 
\multicolumn{1}{l|}{$T_{BO}$}    & \multicolumn{1}{l}{Before Offloading Time} \\ 
\multicolumn{1}{l|}{$T_{Offl}$}  & \multicolumn{1}{l}{Offloading Time}        \\ 
\multicolumn{1}{l|}{$T_{BB}$}    & \multicolumn{1}{l}{Before Boarding Time}   \\ 
\multicolumn{1}{l|}{$T_{Board}$} & \multicolumn{1}{l}{Boarding Time}          \\ 
\multicolumn{1}{l|}{$T_{BPSR}$} & \multicolumn{1}{l}{Before Paxsteps Removed Time}         \\ 
\multicolumn{1}{l|}{$T_{E}$}     & \multicolumn{1}{l}{Exit Time}
\end{tabular}%
\end{table}

\subsubsection{Offloading coordinator}  
\label{sss:agents_OC}
At each step when the \gls{OC} is waiting at its building, the \gls{OC} does the first task that is necessary in the following order: 
\begin{enumerate} 
    \item Send a driver to supply or retrieve pax steps;
    \item Send personnel and \gls{GSE} for civilian offloading;
    \item Send/reserve personnel and \gls{gse} for military offloading;
    \item Check a newly arrived aircraft.
\end{enumerate}

In case no tasks need to be performed it stays in place and repeats the procedure in the next step.

\paragraph{\textit{Pax Steps Property}}   \textcolor{white}{.}  
RETRIEVE: It is assumed that the OC has the knowledge when an aircraft is finished with boarding passengers or if no passengers are needed. It will send a free driver to retrieve the pax steps.\\
SUPPLY: It is assumed that the OC can see the whole tarmac from its office window, but is only aware of the civilian aircraft. When the OC sees a large civilian aircraft arriving, and one driver and one paxsteps are free, the OC will send the driver with paxsteps.

\paragraph{\textit{Arrange Offloading Property}} Resource allocation for offloading is performed chronologically with respect to arrival time. The OC tries to match the offloading needs for each aircraft with the available personnel and \gls{GSE}. When a large aircraft has LD3 ULD it needs at least one civilian handling agents with a highloader and one driver with tug and dollies. When a large aircraft has ICU it needs at least one civilian handling agents with a belt loader and one driver with tug and dollies. When it concerns a small aircraft, then only one driver with tug and dollies is sent to offload the ICU. More explanation about the used function can be found in \ref{ss:property_change}.
The OC assigns the civilian handling agent and/or drivers with their corresponding tasks. After assigning the tasks to its personnel the OC resumes its \textit{waiting} state.

\paragraph{\textit{Military Reservations Property}} After the MOVCON interacts with the OC and requests a highloader reservation, the OC checks the availability, as only one of the highloaders can be used on military aircraft. When available the OC changes the status of the highloader to `reserved' and will not use it.
When the military offloading coordinator interacts with the OC, the OC will do the same as with the highloader reservation. The OC got the time penalty time for the aircraft offloading and boarding from the military offloading coordinator. If all requested \gls{GSE} are available it sets the timer.

\paragraph{\textit{Checking Aircraft Property}} When the OC sees a civilian aircraft arriving, it goes to the aircraft to check the cargo content. In case of a large aircraft it first sends a driver with pax steps ahead. After the OC interacts with the aircraft agent, the OC knows the cargo types and amount, if any. Afterwards, if another aircraft has arrived, the OC will check that aircraft, otherwise it will return to its office. It can check maximum three aircraft with cargo content in one go.\\

\subsubsection{Civilian handling Agents}
The civilian handling agents have two main properties, offloading large civilian aircraft and offloading large military aircraft. In between these two properties the civilian handling agent waits at the office building until instructions are given. The civilian handling agents can also be reserved, which is the same as waiting for the civilian handling agent, but instructions can only be given to offload a military aircraft. The walking speed of the civilian handling agent is 5 $km/h$ and the driving speed, for driving belt loaders and highloaders, is 30 $km/h$.

\paragraph{\textit{Civilian Offloading Property}} First the civilian handling agent walks to its corresponding \gls{GSE}, which is a highloader or a belt loader. Then it drives the \gls{GSE} to its target aircraft together with its assigned driver, who drives the tug and dollies. The duration of the offloading, from one vehicle to another or to the drop off point, is determined by a time penalty. This time penalty takes the type of vehicle and the amount of cargo into account.
When the aircraft is empty, it will communicate this to the driver. Then the civilian handling agent will drive the \gls{GSE} back to its parking spot. After parking the \gls{GSE} the civilian handling agent will walk back to the office.

\paragraph{\textit{Military Offloading Property}} First the civilian handling agent walks to its corresponding \gls{GSE}, which is the reserved highloader. Then it drives the highloader to the target aircraft. There it waits until the time penalty for the complete offloading of a military aircraft is finished. The time penalty is based on the amount of cargo ($kg$) and the offloading rate ($kg/s$). 
Afterwards the civilian handling agent will drive the \gls{GSE} back to its parking spot. After parking the \gls{GSE} the civilian handling agent will walk back to the office.\\

\subsubsection{Drivers}
Drivers can have several tasks, namely: take pax steps to and from aircraft and offloading large and small aircraft. In between these tasks the driver waits at the office building until instructions are given. The walking speed of the driver is 5 $km/h$ and the driving speed is 30 $km/h$. 

\paragraph{\textit{Pax Steps Property}} First the driver walks to the pax steps, assuming the driver knows the location. Then the driver drives the pax steps to a destination point. This destination point is given by the OC during the task instruction. Next, the driver walks back to the office.\\

\paragraph{\textit{Offloading Large Aircraft Property}} The driver walks to the tug and dollies, it then drives the tug and dollies to the target aircraft. The driver waits while the civilian handling agent loads cargo on the dollies. When the dollies are filled, the driver drives to the terminal building to drop the cargo. When it was the last batch of cargo, the civilian handling agent will have communicated this. As long as it is not communicated the driver drives back to the aircraft and the process is repeated. When the driver is finished it drives to the parking spot of the tug and dollies and walks back to the office.

\paragraph{\textit{Offloading Small Aircraft Property}} The driver walks to the tug and dollies, it then drives the tug and dollies to the target aircraft. Subsequently, the driver drives the tug and dollies to the target aircraft. The driver is in contact with the aircraft and therefore always knows how much cargo is left in the aircraft. The driver loads the cargo on the dollies by itself. When the dollies are filled the driver drives to the terminal building to drop the cargo. In case there is more cargo in the aircraft, the driver drives back to the aircraft and repeats the process. When finished, the driver drives to the parking spot of the tug and dollies and walks back to the office.

\subsubsection{ATC} In this model the ATC agent has only one task: to give the information about incoming military flights to the MOVCON. As the ATC agent is a military agent, the agent walks with speed 6 $km/h$.

\paragraph{\textit{Find MOVCON Property}} After a military aircraft calls in, see `Call ATC Property' in Section \ref{sss:aircraft_prop}, the ATC walks to the MOVCON and repeats the given information. The information includes: the aircraft agent ID, arrival time, types of cargo, amount of cargo of each type. After the interaction the ATC walks back to its tower. 

It is assumed a second ATC agent stays in the tower to receive incoming calls. However, that agent is not modelled.\\

\subsubsection{MOVCON} Movement Control is a military unit, in the case study the MOVCON oversees the airport. In this model the MOVCON is the connection between the ATC, the military offloading coordinator and the OC. The MOVCON is a military agent and therefore walks 6 $km/h$. The waiting spot of the MOVCON is the military cargo drop off area. 

If the MOVCON gets information from the ATC about an incoming aircraft with 463L Master Pallets onboard, the MOVCON agent needs to make a reservation, so it could use the civilian \gls{GSE} and civilian handling agent on time of arrival of the aircraft.

\paragraph{\textit{Highloader Reservation Property}} The MOVCON walks to the OC office building. If the OC is not present or available the MOVCON waits. When the OC is present and available, the MOVCON interacts with the OC. The MOVCON gives the arrival time of the aircraft and asks for a highloader reservation. After the interaction the MOVCON walks back to its waiting spot. \\

\subsubsection{Military offloading coordinator} The military offloading coordinator's waiting spot is near the military parking spots, in between the OC's office and the terminal building. The military offloading coordinator is a military agent and walks 6 $km/h$.

\paragraph{\textit{Handle Aircraft Property}} When a military aircraft is parked, the military offloading coordinator walks to the OC office building. If the OC is not present or available, the MOVCON waits. When the OC is present and available the military offloading coordinator interacts with the OC. The military offloading coordinator requests to borrow a tug with dollies. In case of a large military aircraft the military offloading coordinator also requests the civilian handling agent and highloader that have been reserved by the MOVCON. A military handling agent normally drives the requested tug and dollies as well as a forklift truck. However, the military handling agent and forklift trucks are not modelled. Instead, the military offloading coordinator also makes a reservation in its own military handling agent list and forklift truck list. These lists have the same length as the amount of military handling agents or forklift trucks, each entry is either a 0 (= available), an integer (= amount of steps that it is not available) or a string (`reserved'). It is assumed that the military offloading coordinator knows when the reserved \gls{GSE} are available. At that step, the military offloading coordinator changes a `reserved' entry in its military handling agents list and its forklift truck list to a timer entry. After the interaction the military offloading coordinator walks back to its waiting spot. 


\section{Model calibration}
\label{sec:PJIA_prelim_analysis}

In total six parameters needed to be calibrated in comparison to van Liere's model. The calibration was done in two steps, explained in Section \ref{ss:exist_analysis} and Section \ref{ss:new_analysis}. First, the PJIA model was calibrated before implementing the new resource allocation strategy, in order to compare and analyse the PJIA model with respect to van Liere's model. The second step is the calibration after the implementation, to guarantee a realistic output with respect to the expert' knowledge. 

The main performance indicator is TAT, in Section \ref{sss:aircraft_prop} the division of the aircraft TAT is shown. TAT can be split in three main parts: the offloading time, the boarding time and the cumulative waiting time. The parameters to be calibrated are offloading and boarding rates. Therefore only the results for TAT, offloading time and boarding time are shown here. In order to compare the proposed base model with van Liere's model, only the results for large aircraft with cargo and pax are given, as those were discussed in depth in the paper of van Liere.

\subsection{Analysis and comparison to van Liere's model }
\label{ss:exist_analysis}
In Table \ref{tab:01_calibration} the calibration results for the van Liere's model are shown. Here, the same resource allocation strategy is used as in van Liere's model. C0 stands for 0 calibrated input parameters, with respect to van Liere's model.
The results of PJIA model C0 should be the equal to or less than van Liere's model, since some steps are omitted from the new model. In the first row the expectation of experts is shown \cite{weeks,braam,pilot}.

\begin{table}[htbp]
\centering
\caption{First-step calibration results (minutes) for large aircraft with cargo and evacuation.}
\label{tab:01_calibration}
\renewcommand{\arraystretch}{1.1}
\setlength{\tabcolsep}{4.2pt} 
\begin{tabular}{lccc|ccc}
\toprule
\textbf{Source} & \textbf{TAT} & \textbf{Offl.} & \textbf{Board.} & \textbf{TAT} & \textbf{Offl.} & \textbf{Board.} \\
& \multicolumn{3}{c|}{\textbf{Civilian}} & \multicolumn{3}{c}{\textbf{Military}} \\
\midrule
\textit{Experts}        & \textit{120--150} & \textit{60} & \textit{30}   & \textit{60--90}  & \textit{30}   & \textit{30} \\
van Liere model         & 125.2             & 57.5        & 30.6          & 79.8             & 32.5          & 30.2 \\
PJIA model C0           & 129.7             & 53.9        & \textbf{37.3} & \textbf{97.0}    & \textbf{37.0} & \textbf{40.0} \\
PJIA model C3           & 119.1             & 52.5        & \textbf{29.2} & \textbf{79.0}    & \textbf{29.0} & \textbf{30.0} \\
\bottomrule
\end{tabular}
\end{table}

As can be seen in Table \ref{tab:01_calibration} for PJIA model C0, the average offloading time of the civilian aircraft is a bit lower than the experts' value, which is as expected. The civilian boarding time, as well as the military boarding time, are 37.3  and 40.0 minutes respectively, which is more than 20$\%$ higher than they should be. 
The reason for this increase was found to be the assumption of van Liere's boarding rate per pax. This was determined so it matched the experts' boarding time, it was calculated with an average of 30 pax per civilian flight and 60 pax per military flight. However, when looking at the flights in the schedule, the average for civilian flights is 38 pax and for military flights is 80 pax. The reason why the values in van Liere's model matched with the experts is not found. Nevertheless for the PJIA model, the new boarding rates are recalculated with the new actual average pax. These times were close to the experts' times. 
Table \ref{tab:parameters_cal} presents all the parameters new parameters. 
As the boarding times of both the civilian and the military aircraft were too high, this also increased their respective TATs. 

The military offloading time of 37.0 minutes is also too high. The cause for this difference was the combat offload by one of the seven large military aircraft, in van Liere's model one aircraft did a combat offload. This combat offload is included in the average offloading time of van Liere's model. As a combat offload is an exception and therefore not part of the general offloading time, the calibration of the military offloading time in van Liere's model was not correct. Especially as there are only 7 large military aircraft with cargo and pax, one combat offload has a large influence. As a result new military 463L cargo offload rate, from aircraft to GSE, was implemented. 

To summarise, three parameters are calibrated, the 463L cargo offloading rate from aircraft to GSE, the military boarding rate and the civilian boarding rate, as seen in Table \ref{tab:parameters_cal}. These three changes give the results on the last row of Table \ref{tab:01_calibration}, where PJIA model C3 is given. The calibrated parameters have the desired effect, although the civilian TAT and offloading time are low. No immediate reason was found for this, except that some steps were eliminated. As a new offloading strategy will be implemented in the next part, the civilian TAT and offloading time will be analysed there. 


\begin{table}[htbp]
\centering
\caption{Parameters changed for calibration in the first step.}
\label{tab:parameters_cal}
\renewcommand{\arraystretch}{1.15}
\setlength{\tabcolsep}{3.5pt}
\begin{tabular}{p{3.5cm}ccc}
\toprule
\textbf{Parameter} & \textbf{van Liere (C0)} & \textbf{PJIA (C6)} & \textbf{Change ID} \\
\midrule
\multicolumn{4}{l}{\cellcolor[HTML]{DAE8FC}\textbf{Offloading Rates}} \\
463L (AC to GSE) [kg/s]     & 7.5  & 10.0  & C1 \\
ICU (general) [kg/s]        & 5.0  & 4.5   & C4 \\
ULD (AC to GSE) [kg/s]      & 7.1  & 5.7   & C5 \\
ULD (GSE to TB) [kg/s]      & 45.0 & 40.0  & C6 \\
\addlinespace[0.5ex]
\multicolumn{4}{l}{\cellcolor[HTML]{DAE8FC}\textbf{Boarding Rates}} \\
Boarding (military) [s/pax] & 30.0 & 22.5  & C2 \\
Boarding (civilian) [s/pax] & 60.0 & 47.0  & C3 \\
\bottomrule
\end{tabular}
\end{table}

\subsection{Calibrating base model}
\label{ss:new_analysis}

In this section, the PJIA model with new strategy is calibrated to fit the expert knowledge.
The base model starts off with 3 changed parameter compared to van Liere's model. From this point on the PJIA model uses the \textit{`maximum available'} resource allocation strategy. 

\begin{table}[htbp]
\centering
\caption{Second-step calibration results (minutes) for large aircraft, comparing PJIA models with expert values.}
\label{tab:11_calibration}
\renewcommand{\arraystretch}{1.15}
\setlength{\tabcolsep}{4pt}
\begin{tabular}{lccc|ccc}
\toprule
\textbf{Model} & \textbf{TAT} & \textbf{Offl.} & \textbf{Board.} & \textbf{TAT} & \textbf{Offl.} & \textbf{Board.} \\
              & \multicolumn{3}{c|}{\textbf{Civilian}} & \multicolumn{3}{c}{\textbf{Military}} \\
\midrule
\textit{Experts}     & \textit{120--150} & \textit{60}   & \textit{30}   & \textit{60--90}  & \textit{30}  & \textit{30} \\
PJIA model C3        & \textbf{115.7}    & \textbf{48.1} & 29.2          & 79.0             & 29.0         & 30.0 \\
PJIA model C6        & \textbf{126.6}    & \textbf{58.8} & 29.0          & 79.0             & 29.0         & 30.0 \\
\bottomrule
\end{tabular}
\end{table}

In Table \ref{tab:11_calibration}, it is shown that because of the new offloading strategy, the average offloading time is more than 4 minutes faster than before. This also causes the average TAT to drop. As these numbers are unexpected according to experts, some parameters need to be recalibrated. In Table \ref{tab:parameters_cal} the changed parameters are shown: the offloading rate of ULD from aircraft to GSE, and from GSE to terminal building are lowered, the ICU offloading rate is also lowered. This gives the outputs of PJIA model C6, where all values are in between the expert values.

van Liere's model and PJIA model have exactly the same amount of military aircraft, with exactly the same arrival time. In both models there are 7 large civilian aircraft. Their arrival times were randomized in van Liere's model, meaning that in each of the 120 schedules it is randomized as well. In van Liere's model there were on average 30 small aircraft with cargo and 15 small aircraft without, the schedules of the base model have on average 30.3 and 15.8 respectively.

\section{Experiments}
\label{sec:methodology}
This section presents the methodology to answer the main research question, namely what is the effect of unannounced incoming flights on the performance of the system using different resource allocation strategies. In Section \ref{ss:meth_experiments} the test cases will be explained. Afterwards, in Section \ref{ss:property_change}, the implementation of the test cases in the model is elaborated.

\subsection{Experiments: strategies and scenarios}
\label{ss:meth_experiments}
In a situation after a natural disaster event most scheduled flights fly if possible. However, extra flights come in and their arrival is not always announced. According to experts, this causes a disruption in the planning \cite{weeks}. 
An overview of all test cases in this research are summarised in Table \ref{tab:experiments}. 

\begin{table*}[htbp]
\centering
\caption{Simulation test cases.}
\label{tab:experiments}
\renewcommand{\arraystretch}{1.25}
\setlength{\tabcolsep}{5pt}
\begin{tabular}{llc|ccc}
\toprule
& & & \multicolumn{3}{c}{\cellcolor[HTML]{DAE8FC}\textbf{Scenarios}} \\
\cmidrule(lr){4-6}
& \textbf{Strategy} & 
\textbf{} & 
\cellcolor[HTML]{EFEFEF}A: Unknown Flight Schedule & 
\cellcolor[HTML]{EFEFEF}B: Known Flight Schedule & 
\cellcolor[HTML]{EFEFEF}C: Incomplete Flight Schedule \\
\midrule
\multirow{1}{*}{}
& Max available & 
\textbf{1} & 
Case 1A & Case 1B & Case 1C u1, u7 \\
\multirow{1}{*}{}
& Max available + anticipation & 
\textbf{2} & 
Case 2A & Case 2B & Case 2C u1, u7 \\
\bottomrule
\end{tabular}
\end{table*}

\subsubsection{Incoming Flight scenarios} 
In the case study the airport was not operational so most flights were not scheduled. However, the military got the information on incoming flights in the morning of each day, the civilian flights all arrived unannounced. According to experts, in general the civilian schedule is known, but there are always unannounced aircraft causing delays \cite{weeks}.
To achieve a better understanding of the influence of incomplete knowledge of incoming flights, three different incoming civilian flight scenarios are created:
\begin{enumerate}[label=\Alph*:]    
    \item \textbf{Unknown Flight Schedule}: The flight schedule is not known by the OC: this is what happened in the case study and considered in van Liere's model. 
    \item \textbf{Known Flight Schedule}: The \gls{OC} has full knowledge of the incoming civilian flight schedule. This scenario is used to compare the case with complete information with cases with incomplete information about the flight schedule.
    \item \textbf{Incomplete Flight Schedule}: The \gls{OC} has access to the incoming civilian flight schedule, but unannounced aircraft will come in. Implementation is done by making some large aircraft that are known in the second scenario unknown. 
    This scenario is implemented to analyse how unannounced flights can disrupt the overall system.
\end{enumerate}
In Table \ref{tab:experiments} it can be seen that for scenario C there are two test cases, \textit{u1} and \textit{u7}. This stands for one unannounced large civilian aircraft and seven unannounced large civilian aircraft respectively. There are only seven large civilian aircraft per schedule in the system, meaning the first is chosen to test the effects of one unannounced aircraft on the performance of the system and the other to test the effects when all large aircraft come in unannounced.

To be able to compare all the experiments, the same schedules should be compared. 
In van Liere's model the military flight schedule was fixed, this schedule is maintained. The civilian aircraft came in randomly, when there were less than 4 civilian aircraft on the tarmac, a new aircraft came in. Size and content of aircraft was randomized as well.
As the exact schedule of the case study is not known, one schedule is not sufficient to perform the research. Therefore the base model was run in the same way as the van Liere model, for 120 times. The incoming flight schedule of this run is kept as a base, all other test cases were executed with the same 120 schedules.\\

\subsubsection{Resource Allocation Strategies} 
To test these scenarios, the model needs realistic offloading strategies. In Section \ref{sss:CS_assump_lim}, the realistic offloading, or resource allocation, strategies, as established by experts, are given. In the base model, chronological offloading and maximum available resources are already implemented. Having knowledge of the schedule is not used in this strategy, thus to make the model more realistic and able to react to unexpected circumstances a cognitive function needs to be implemented in the \gls{OC} agent. This cognitive function needs to make the \gls{OC} agent able to use the information at hand to make a prediction and based on that prediction a decision on resource distribution. As there was no electricity and thus no computers, all calculations in the model should be reasonable for a human to perform manually and quickly. 
The resource allocation strategies are summarised as:
\begin{enumerate}[label=\arabic*:]
    \item \textbf{Maximal available}: One civilian handling agent with GSE and one or two driver(s) with a tug with dollies per aircraft, per cargo door. The amount that is sent depends on the amount that is available at that time step. Even if there are two cargo doors, only one civilian handling agent with GSE can be sent.
    \item \textbf{Maximal set available + anticipation}: Is the same as the previous, only now the OC is also able to estimate the length of a task and can use that information to estimate when personnel and GSE will be free, as well as to take the needs for the next aircraft into account. 
\end{enumerate}

\begin{table*}[htbp]
\centering
\caption{Resource sets for each cargo type, for each aircraft type and strategy.}
\label{tab:res_sets}
\renewcommand{\arraystretch}{1.25}
\setlength{\tabcolsep}{5pt}
\begin{tabular}{ccccccc}
\rowcolor[HTML]{ECF4FF}
\textbf{Cargo Type} & \textit{AC Type} & \textit{Strategy} &
\begin{tabular}[c]{@{}c@{}}\textbf{Civilian} \\ \textbf{handling agent}\end{tabular} &
\textbf{Highloader} & \textbf{Belt loader} &
\begin{tabular}[c]{@{}c@{}}\textbf{Driver +} \\ \textbf{Tug and dollies}\end{tabular} \\
\midrule
\cellcolor[HTML]{EFEFEF}\textbf{ULD} 
& \textit{Civ Large} & \textit{1} & 1 & 1 & - & 1 \\
\cellcolor[HTML]{EFEFEF} 
& \textit{Civ Large} & \textit{2} & 1 & 1 & - & 1 or 2 \\
\midrule
\cellcolor[HTML]{EFEFEF}\textbf{ICU} 
& \textit{Civ Large} & \textit{1}     & 1 & - & 1 & 1 \\
\cellcolor[HTML]{EFEFEF} 
& \textit{Civ Large} & \textit{2}     & 1 & - & 1 & 1 or 2 \\
\cellcolor[HTML]{EFEFEF} 
& \textit{Civ Small} & \textit{1 \& 2} & - & - & - & 1 \\
\midrule
\cellcolor[HTML]{EFEFEF}\textbf{463L} 
& \textit{Mil Large} & \textit{1 \& 2} & 1 & 1 & - & - \\
\bottomrule
\end{tabular}
\end{table*}

In Table \ref{tab:res_sets} the amount that can be sent per cargo type, aircraft type and strategy is shown.

\SetKwInput{KwInput}{Input}                
\SetKwInput{KwOutput}{Output}              
\SetKwInput{KwFunction}{Function}

\begin{algorithm}
\DontPrintSemicolon

\tcp{RA\_function returns Option Set containing the Amount Assigned, Estimated ULD and ICU Offloading Step, and the step where offloading can start: \{AMA, est\_ULD, est\_ICU, start\_step\}, for args see Table~\ref{tab:RA_functions_options}}
\KwFunction{RA\_function(args)}{
    \textbf{return} Option Set
}

\BlankLine
\tcp{Start of algorithm}
\textbf{Algorithm: Generate Best Option}

\tcp{List of estimated resources, ID of aircraft A, ID of aircraft B}
\KwInput{est\_rss, A\_ID, B\_ID}
\KwOutput{Best Option}

\BlankLine
\tcp{Combination of arguments of the RA\_function(), see Table~\ref{tab:RA_functions_options}}
\ForEach{Possible Argument Combination}{
    
    Option Set A $\gets$ RA\_function()\;
    
    \tcp{Update the list of estimated resources to include Option Set A}
    est\_rss $\gets$ rss\_update(est\_rss, Option Set A)\;
    
    \tcp{Generate Option Set B with new est\_rss}
    Option Set B $\gets$ RA\_function()\;
    
    \tcp{Maximum of the estimated ULD and ICU offloading finish steps of Option Set A and B}
    LatestEndTime(Option Set A) $\gets$ max(est\_ULD of A, est\_ICU of A, est\_ULD of B, est\_ICU of B)\;
    
    \tcp{Aircraft should not wait longer than max start time}
    \If{StartTime(Option Set A) $>$ MaxStartTime(Option Set A)}{
        LatestEndTime(Option Set A) $\gets$ LatestEndTime(Option Set A) + 100000\;
    }
}

Best Option $\gets$ Option Set A with earliest Latest End Time\;

\caption{Generate Best Option}
\end{algorithm}

\subsection{Implementation of strategies and scenarios}
\label{ss:property_change}
In this section, the properties and functions that change for strategy 2 and scenario B and C, are explained. \\

\subsubsection{Including anticipation}
For small aircraft there is only one option for the resource allocation. Hence, nothing will change for strategy 2. For large aircraft, a function is implemented to find the best resource allocation plan, in terms of total TAT, for aircraft A being serviced, taking the next aircraft B into account. 

It is important to note that the OC only goes into the \textit{`resource$\_$allocation'} state, when there are enough resources to start offloading at least one of the cargo types of aircraft A. 

In Algorithm 1, the algorithm finding the best resource allocation option is shown. The function \texttt{RA\_function()} in  the algorithm is given to indicate either the \texttt{RA$\_$available()} function or the  \texttt{RA$\_$ahead()} function. The first is used to determine the maximum \textit{Option Set} for the current time step, while the second is used to determine the minimum or maximum \textit{Option Set} and establish when these \textit{Option Sets} are possible. The output for both functions is the same, however the input slightly changes, as can be seen in Table \ref{tab:RA_functions}.

This function uses the \texttt{RA$\_$available()} of strategy 1, as well the \texttt{RA$\_$ahead()} function, which is used to determine future availability.
Both the \texttt{RA$\_$available()} and \texttt{RA$\_$ahead()} return the \textit{\textit{Option Set}} containing the following:

\begin{itemize}
    \item \textit{AMA}: Amount Assigned
    \item \textit{est$\_$ULD}:
    Estimated ULD offloading finish step
    \item \textit{est$\_$ICU}:
    Estimated ICU offloading finish step
    \item \textit{start$\_$step}: (Estimated) step when offloading can/will start  \\
\end{itemize}

\begin{table*}[htbp]
\centering
\caption{Elaboration on the \texttt{RA\_functions}: \texttt{RA\_available()} and \texttt{RA\_ahead()}, with a brief explanation, return, and input.}
\label{tab:RA_functions}
\renewcommand{\arraystretch}{1.3}
\setlength{\tabcolsep}{8pt}
\begin{tabular}{p{2.5cm} p{12.5cm}}
\toprule
\multicolumn{2}{l}{\textbf{\texttt{RA\_available()}}} \\
\midrule
What:   & Maximum Option Set for available resources \\
Return: & \textit{Options Set} = \{\textit{AMA, est\_ULD, est\_ICU, start\_step}\} \\
Input:  & 
\textit{free\_rss}: Free resources list \newline
\textit{X\_ID}: ID of aircraft X \\
\midrule
\multicolumn{2}{l}{\textbf{\texttt{RA\_ahead()}}} \\
\midrule
What:   & Minimum and/or maximum Option Set for one or both cargo types \\
Return: & \textit{Options Set} = \{\textit{AMA, est\_ULD, est\_ICU, start\_step}\} \\
Input:  &
\textit{est\_rss}: resources estimation list \newline
\textit{X\_ID}: ID of aircraft X \newline
\textit{`ULD', `ICU', `All'}: choose which cargo type \newline
\textit{`Max', `Min', `Best'}: choose which Option Set \\
\bottomrule
\end{tabular}
\end{table*}

The \textit{Best Option} algorithm requires the list of estimated resources as input. This list contains each resource and the estimated timestep when the task is finished. In this list a distinction is made between available or free resources and unavailable resources. For the unavailable resources, the OC knows the estimated timestep when that resource is free. Here, the estimations for each task are the average times the resources are busy with a certain task, taken from case 1A. For the offloading times the OC has a different estimation for each cargo type, per amount of cargo and per amount of drivers. The use of two drivers can reduce the offloading time between 11-13$\%$ for ULD cargo and 32-35$\%$ for ICU cargo. The range is dependent on the amount of cargo. An overview of these times is given in Table \ref{tab:est_offl}. Furthermore, the aircraft IDs of the two first aircraft to be offloaded are given. The IDs are linked to the internal knowledge the OC has about that aircraft. Finally, the algorithm outputs the best option, which is then used to determine the next state and/or task of the OC. If the best option has a \textit{start$\_$step} equal to the current timestep, the OC's new state is \textit{`interacting$\_$personnel'}, else the OC sets its state to \textit{`waiting'}.

\begin{table}[h!]
\centering
\caption{Average offloading times in minutes per cargo type, per amount of cargo and per amount of drivers.}
\label{tab:est_offl}
\begin{tabular}{ll|lll}
             & Cargo Amount ($kg$)       & 15000 & 18000 & 20000 \\ \hline
\textbf{ULD} & 1 driver  & 52  & 61  & 70  \\
             & 2 drivers & 46  & 54  & 61  \\ \hline 
             & Cargo Amount ($kg$)        & 5000  & 6000  & 8000  \\ \hline
\textbf{ICU} & 1 driver  & 35  & 37  & 54  \\
             & 2 drivers & 21  & 25  & 33
\end{tabular}
\end{table}


There are two sets of possible \texttt{RA$\_$function()} combinations for aircraft A and B. These combinations are given in Table \ref{tab:RA_functions_options}. When the minimum set of resources for aircraft A cannot be reached at the current time step, it means that there are enough resources for only one cargo type to be offloaded. In that case, the option set for aircraft B is equal to the option set for aircraft A, but with the other cargo type. For example, the first option set that will be generated uses for \textit{Option Set A}, aircraft A with only ULD cargo. For \textit{Option Set B}, it is aircraft A with ICU cargo. In case the minimum set of resources for aircraft A can be reached at that time step, the minimum, the maximum, and the maximum available \textit{Option Sets} are evaluated.\\

\begin{table*}[htbp]
\centering
\caption{All resource allocation function combinations between Option Set A and Option Set B.}
\label{tab:RA_functions_options}
\renewcommand{\arraystretch}{1.3}
\setlength{\tabcolsep}{6pt}
\begin{tabular}{p{3.5cm} c p{5.5cm} p{5.5cm}}
\toprule
& \textit{Set i} & \textbf{Option Set A} & \textbf{Option Set B} \\
\midrule
\multirow{4}{*}{\texttt{Time\_min\_A > current step}} 
& \textit{1:} & \texttt{RA\_ahead(A\_ID, est\_rss, ULD, min)} & \texttt{RA\_ahead(A\_ID, est\_rss, ICU, Best)} \\
& \textit{2:} & \texttt{RA\_ahead(A\_ID, est\_rss, ULD, max)} & \texttt{RA\_ahead(A\_ID, est\_rss, ICU, Best)} \\
& \textit{3:} & \texttt{RA\_ahead(A\_ID, est\_rss, ICU, min)} & \texttt{RA\_ahead(A\_ID, est\_rss, ULD, Best)} \\
& \textit{4:} & \texttt{RA\_ahead(A\_ID, est\_rss, ICU, min)} & \texttt{RA\_ahead(A\_ID, est\_rss, ULD, Best)} \\
\midrule
\multirow{3}{*}{\texttt{Time\_min\_A ≤ current step}} 
& \textit{1:} & \texttt{RA\_ahead(A\_ID, est\_rss, All, min)} & \texttt{RA\_ahead(B\_ID, est\_rss, All, Best)} \\
& \textit{2:} & \texttt{RA\_ahead(A\_ID, est\_rss, All, max)} & \texttt{RA\_ahead(B\_ID, est\_rss, All, Best)} \\
& \textit{3:} & \texttt{RA\_available(A\_ID, free\_rss)}      & \texttt{RA\_ahead(B\_ID, est\_rss, All, Best)} \\
\bottomrule
\end{tabular}
\end{table*}

\section{Results}
\label{sec:Results}
In this section the results of the test cases are presented. The main results are considered in Section \ref{ss:res-stratscen}. In Section \ref{ss:res-atif} the results of the test cases with different Arrival Time Interval Factors (ATIF) are specified.

\begin{table}[htbp]
\centering
\caption{The average TAT and Offloading Times, including standard deviation, for large civilian aircraft per test case.}
\label{tab:av-tat-offl}
\renewcommand{\arraystretch}{1.2}
\setlength{\tabcolsep}{3pt}
\begin{tabular}{clcccc}
\toprule
\multicolumn{2}{c}{} & \multicolumn{4}{c}{\cellcolor[HTML]{DAE8FC}\textbf{Scenarios}} \\
\cmidrule(lr){3-6}
\rowcolor[HTML]{DAE8FC}
\textbf{Strategy} & \cellcolor[HTML]{EFEFEF}\textbf{Metric}
& \cellcolor[HTML]{EFEFEF}A 
& \cellcolor[HTML]{EFEFEF}B 
& \cellcolor[HTML]{EFEFEF}C u1 
& \cellcolor[HTML]{EFEFEF}C u7 \\
\rowcolor[HTML]{DAE8FC}
\textbf{1} & TAT   & 130.4 $\pm$ 5.6 & 110.8 $\pm$ 4.8 & 111.8 $\pm$ 4.7 & 121.4 $\pm$ 5.2 \\
\rowcolor[HTML]{DAE8FC}
          & Offl  & 60.0  $\pm$ 3.9 & 59.4  $\pm$ 3.8 & 59.0  $\pm$ 3.7 & 59.0  $\pm$ 3.9 \\
\rowcolor[HTML]{DAE8FC}
\textbf{2} & TAT   & 128.7 $\pm$ 5.9 & 110.3 $\pm$ 4.8 & 111.2 $\pm$ 4.7 & 120.5 $\pm$ 5.4 \\
\rowcolor[HTML]{DAE8FC}
          & Offl  & 58.2  $\pm$ 3.7 & 58.2  $\pm$ 3.6 & 57.8  $\pm$ 3.7 & 57.9  $\pm$ 3.7 \\
\bottomrule
\end{tabular}
\end{table}

\subsection{Comparison of strategies and scenarios}
\label{ss:res-stratscen}
The average TAT and offloading time per test case are presented in Table \ref{tab:av-tat-offl}. The average offloading times do not vary significantly per test case, but the average TAT changes per scenario. The average boarding time is the same for all cases, namely 29.1 minutes with a standard deviation of 1.1 minute.  

The Cliff's-delta test is a non-parametric test to test the effect size of two datasets. If the effect size is large it is very likely that a datapoint of set 1 is smaller than a datapoint of set 2. The thresholds for each magnitude used in this research (negligible, small, medium, large) are defined by Hess and Kromrey \cite{effect_size}. 
As the input of all test cases is the same, meaning the flight schedule and agents, but the strategy or scenario is different, this test can also be used to see if one test case performs differently than another.

\begin{table*}[htbp]
\centering
\caption{Compare results per scenario with Cliff's delta test (N = negligible).}
\label{tab:cliff-delta-scen}
\renewcommand{\arraystretch}{1.2}
\setlength{\tabcolsep}{6pt}
\begin{tabular}{c||c|cccccc}
\rowcolor[HTML]{DAE8FC} 
\textbf{Compare} & \textbf{TAT} & \textbf{T$_{BPS}$} & \textbf{T$_{BC}$} & \textbf{T$_{BO}$} & \textbf{T$_{Offl}$} & \textbf{T$_{Board}$} & \textbf{T$_{BPSR}$} \\
\hline \hline
\rowcolor[HTML]{EFEFEF} 
1A vs. 1B       & \textbf{Large}  & Large  & Large & Large & N & N & Medium \\
\rowcolor[HTML]{FFFFFF} 
1B vs. 1C u1    & \textbf{N}      & N      & N     & N     & N & N & N      \\
\rowcolor[HTML]{EFEFEF} 
1C u1 vs. 1C u7 & \textbf{Medium} & Large  & Large & Large & N & N & N      \\
\rowcolor[HTML]{FFFFFF} 
1C u7 vs. 1A    & \textbf{Small}  & Medium & Small & Small & N & N & Medium \\
\hline
\rowcolor[HTML]{EFEFEF} 
2A vs. 2B       & \textbf{Large}  & Large  & Large & Large & N & N & Medium \\
\rowcolor[HTML]{FFFFFF} 
2B vs. 2C u1    & \textbf{N}      & N      & N     & N     & N & N & N      \\
\rowcolor[HTML]{EFEFEF} 
2C u1 vs. 2C u7 & \textbf{Small}  & Large  & Large & Large & N & N & N      \\
\rowcolor[HTML]{FFFFFF} 
2C u7 vs. 2A    & \textbf{Small}  & Medium & Small & Small & N & N & Medium \\
\bottomrule
\end{tabular}
\end{table*}

\begin{figure}[!tbp]
  \centering
  \begin{minipage}[b]{0.53\textwidth}
    \includegraphics[width=\textwidth]{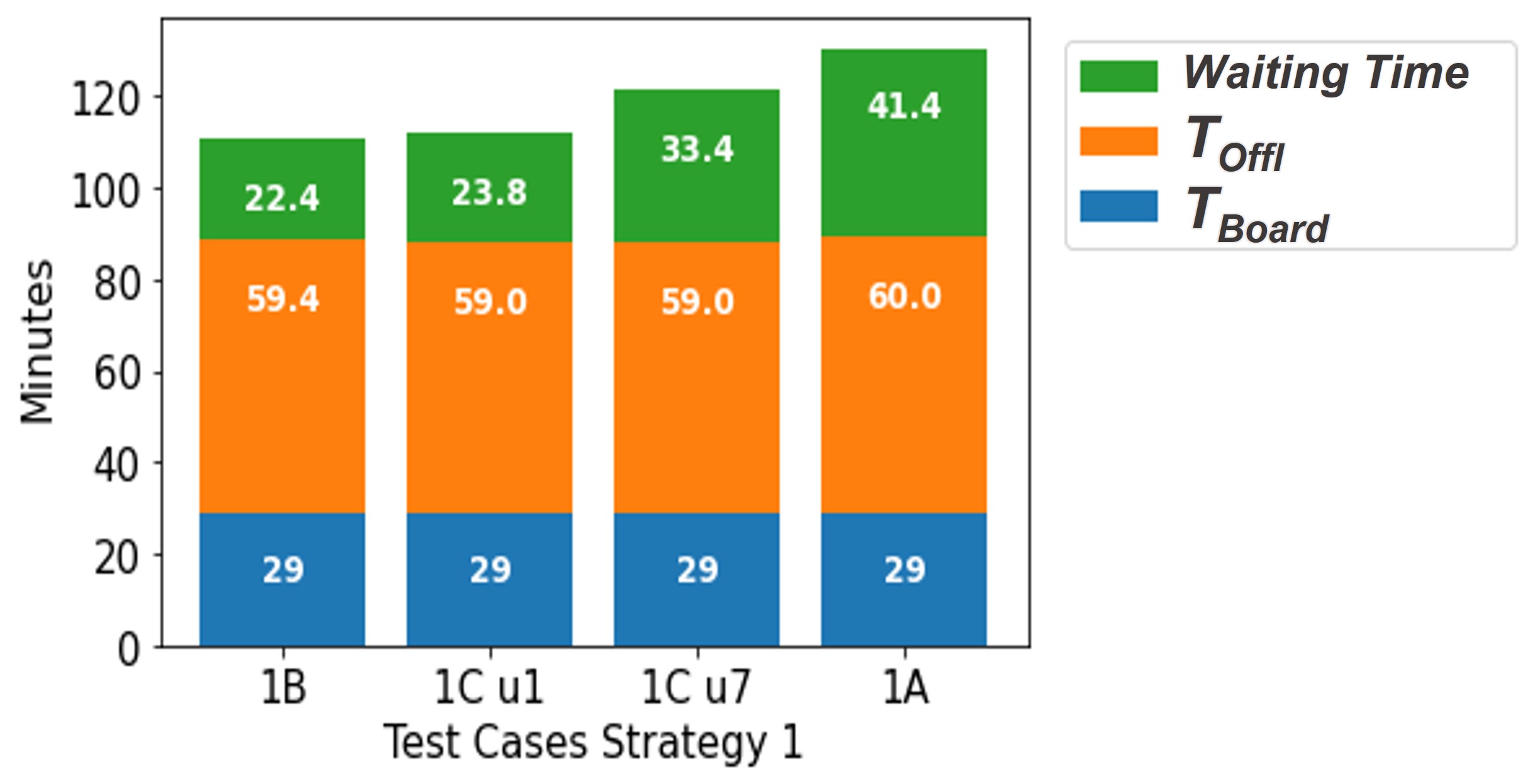}
    \caption{Distribution of the Turnaround Time of large civilian aircraft for all test cases of strategy 1 in order of decreasing knowledge.}
    \label{fig:barchart_distrib_TAT}
  \end{minipage}
  \hfill
  \begin{minipage}[b]{0.46\textwidth}
    \includegraphics[width=\textwidth]{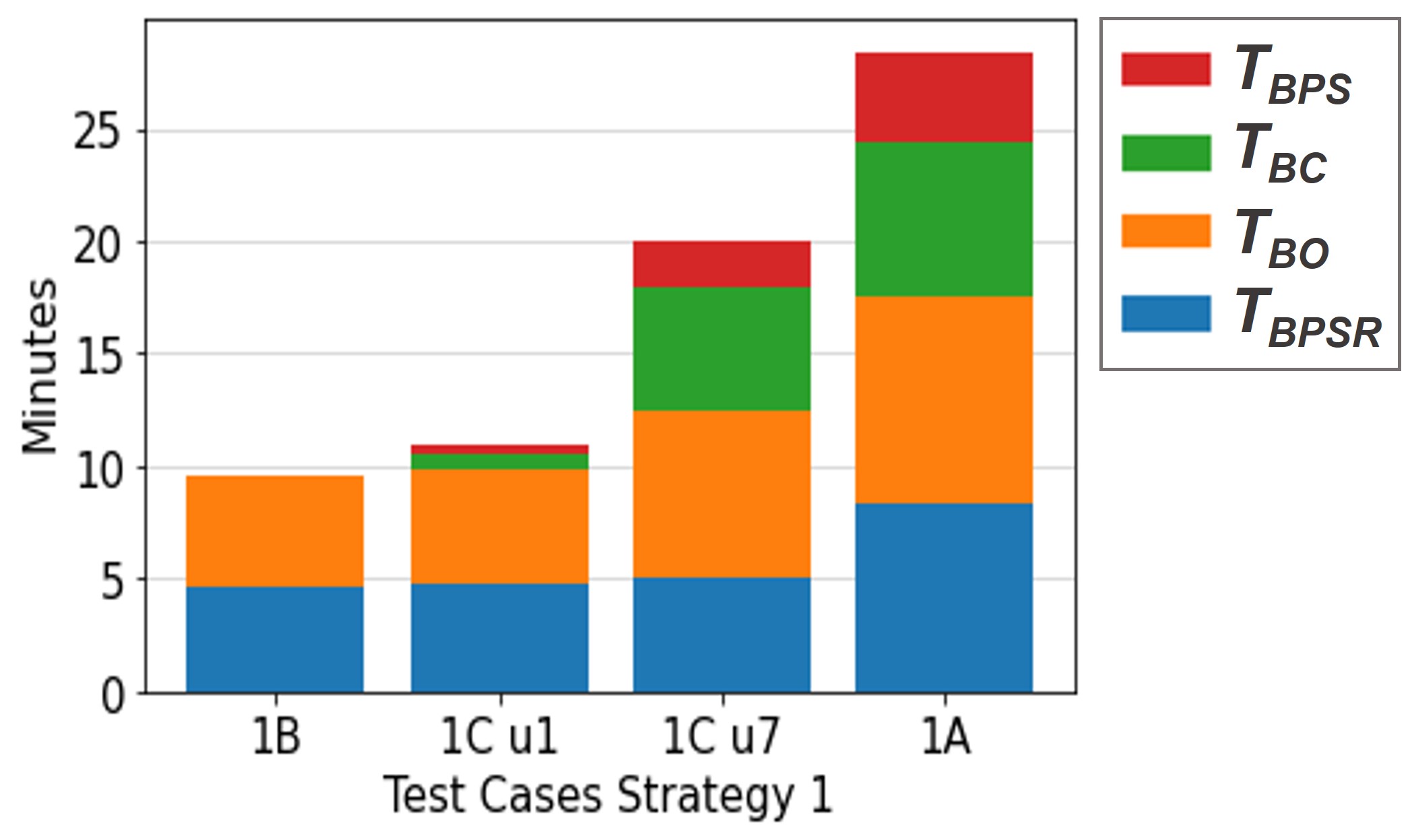}
    \caption{Distribution of the Waiting Times of large civilian aircraft for all test cases of strategy 1 in order of decreasing knowledge.}
    \label{fig:barchart_distrib_WT}
  \end{minipage}
\end{figure}

The two strategies were compared for each scenario using the Cliff's-delta test. According to that test, the differences are all negligible. When comparing the scenarios for each strategy, see Table \ref{tab:cliff-delta-scen}, small to large differences can be detected. One unannounced aircraft does not influence the performance of the system in either of the strategies, whereas seven unannounced aircraft do make a difference. This difference in TAT is mainly due to the change in waiting times as the differences in offloading and boarding times are all negligible.
This is also shown in Figure \ref{fig:barchart_distrib_TAT}, which shows the distribution of the TAT for strategy 1 per scenario in order of decreasing available knowledge. From this figure it is clear that as expected, the boarding times, that are computed as a time penalty, do not change. Furthermore, the offloading times barely vary per test case. Only the total waiting times increase with decreasing available information. 

Figure \ref{fig:barchart_distrib_WT} presents the different waiting times.
\textit{T$_{BPSR}$} does not vary significantly, unless the flight schedule is fully unknown. The \textit{T$_{BO}$} increases slightly with less available information. When the schedule is known, the cargo content does not need to be checked, hence the \textit{T$_{BC}$} increases with increasing unannounced flights. Interestingly, scenario A still has higher \textit{T$_{BC}$} even though there are only seven large civilian aircraft for each run. Next to that, the \textit{T$_{BPS}$} also increases with decreasing knowledge. 
 
From these results, one can see that the offloading times do not change, regardless of the strategy or scenario. Additionally, the overall difference between strategies is negligible. This might be because the resources are always sufficient to serve the demand. The arrival time interval, representing the demand for resources, will be changed to investigate this phenomenon further.

\subsection{Effects of change in arrival time interval}
\label{ss:res-atif}
To be able to compare the experiments with a change in arrival time interval, all schedules were run with an Arrival Time Interval Factor (ATIF), which modifies the arrival times.

\begin{table*}[htbp]
\centering
\caption{Overview of average TAT and Offloading times per ATIF for all cases.}
\label{tab:av-tat-off-atif}
\renewcommand{\arraystretch}{1.2}
\setlength{\tabcolsep}{4pt}
\begin{tabular}{r|cc|cc||cc|cc}
\toprule
\multirow{2}{*}{\textit{ATIF}} 
& \multicolumn{2}{c|}{\textbf{TAT}} 
& \multicolumn{2}{c||}{\textbf{Offloading}} 
& \multicolumn{2}{c|}{\textbf{TAT}} 
& \multicolumn{2}{c}{\textbf{Offloading}} \\
& \multicolumn{4}{c||}{\textbf{Case 1A}} 
& \multicolumn{4}{c}{\textbf{Case 2A}} \\
\midrule
1.2 & 125.3 & $\pm$ 5.2 & 59.2 & $\pm$ 3.8 & 123.6 & $\pm$ 5.6 & 57.1 & $\pm$ 3.7 \\
1   & 130.4 & $\pm$ 5.6 & 60.0 & $\pm$ 3.9 & 128.7 & $\pm$ 5.9 & 58.2 & $\pm$ 3.7 \\
0.8 & 140.4 & $\pm$ 8.2 & 62.4 & $\pm$ 5.6 & 139.7 & $\pm$ 7.3 & 61.1 & $\pm$ 4.6 \\
0.6 & 162.1 & $\pm$ 11.2 & 65.0 & $\pm$ 5.1 & 161.6 & $\pm$ 13.3 & 64.8 & $\pm$ 6.3 \\
\midrule
& \multicolumn{4}{c||}{\textbf{Case 1B}} & \multicolumn{4}{c}{\textbf{Case 2B}} \\
\midrule
1.2 & 109.4 & $\pm$ 4.7 & 58.3 & $\pm$ 3.7 & 108.7 & $\pm$ 4.7 & 57.0 & $\pm$ 3.6 \\
1   & 110.8 & $\pm$ 4.8 & 59.4 & $\pm$ 3.8 & 110.3 & $\pm$ 4.8 & 58.2 & $\pm$ 3.6 \\
0.8 & 113.2 & $\pm$ 5.4 & 60.6 & $\pm$ 4.0 & 113.5 & $\pm$ 5.4 & 59.5 & $\pm$ 4.0 \\
0.6 & 120.1 & $\pm$ 6.8 & 62.9 & $\pm$ 4.6 & 121.0 & $\pm$ 6.5 & 62.8 & $\pm$ 4.6 \\
\midrule
& \multicolumn{4}{c||}{\textbf{Case 1C u1}} & \multicolumn{4}{c}{\textbf{Case 2C u1}} \\
\midrule
1.2 & 110.3 & $\pm$ 4.6 & 58.1 & $\pm$ 3.8 & 109.6 & $\pm$ 4.9 & 57.0 & $\pm$ 3.8 \\
1   & 111.8 & $\pm$ 4.7 & 59.0 & $\pm$ 3.7 & 111.2 & $\pm$ 4.7 & 57.8 & $\pm$ 3.7 \\
0.8 & 114.2 & $\pm$ 5.1 & 60.1 & $\pm$ 4.1 & 114.0 & $\pm$ 5.3 & 59.1 & $\pm$ 4.0 \\
0.6 & 121.3 & $\pm$ 6.9 & 62.9 & $\pm$ 4.6 & 121.8 & $\pm$ 6.4 & 62.4 & $\pm$ 4.3 \\
\midrule
& \multicolumn{4}{c||}{\textbf{Case 1C u7}} & \multicolumn{4}{c}{\textbf{Case 2C u7}} \\
\midrule
1.2 & 119.2 & $\pm$ 4.7 & 58.5 & $\pm$ 3.6 & 118.1 & $\pm$ 4.8 & 57.1 & $\pm$ 3.8 \\
1   & 121.4 & $\pm$ 5.2 & 59.0 & $\pm$ 3.9 & 120.5 & $\pm$ 5.4 & 57.9 & $\pm$ 3.7 \\
0.8 & 125.5 & $\pm$ 6.0 & 60.6 & $\pm$ 3.9 & 124.8 & $\pm$ 6.3 & 59.3 & $\pm$ 4.1 \\
0.6 & 134.4 & $\pm$ 8.2 & 62.6 & $\pm$ 4.7 & 134.5 & $\pm$ 8.6 & 62.0 & $\pm$ 4.9 \\
\bottomrule
\end{tabular}
\end{table*}

In Table \ref{tab:av-tat-off-atif} an overview is given of all the average TAT and Offloading times for all cases with four different ATIFs: 1.2, 1.0, 0.8, 0.6. An increment and decrement of 0.2 is chosen to systematically observe an effect on the system's performance. Note that factor 1.0 is the original arrival time interval.

Comparing the strategies, the TAT and $T_{Offl}$ stay the same. There is a larger difference between no knowledge of incoming aircraft and seven unannounced aircraft, in comparison to ATIF 1. 

Moreover, it can be seen in Table \ref{tab:av-tat-off-atif} that, as expected, both the TAT and offloading time increase with decreasing ATIF. The difference between strategy 1 and strategy 2 is still negligible, which is confirmed by the Cliff's-delta test. Even though the offloading times increase with smaller ATIF, this increase is not as substantial as the increase of the TATs. This suggests that it is again mainly the waiting times that influence the increase in TATs.
Table \ref{tab:av-tat-off-atif} also shows that the standard deviation increases with decreasing interval.

\begin{table}[]
\centering
\caption{Cliff's deltas for the TAT and offloading time for ATIF 0.6 and ATIF 1.2, when comparing strategies.}
\label{tab:deltas_strat_atif}
\resizebox{8.5cm}{!}{%
\begin{tabular}{c|cc|cc|}
\cline{2-5}
\multicolumn{1}{l|}{} & \multicolumn{2}{c|}{\cellcolor[HTML]{DAEEF3}delta TATs} & \multicolumn{2}{c|}{\cellcolor[HTML]{DAEEF3}delta $T_{Offl}$} \\ \cline{2-5} 
                                      & ATIF 0.6 & ATIF 1.2 & ATIF 0.6 & ATIF 1.2 \\ \hline
\multicolumn{1}{|c|}{1A vs. 2A}       & 0.004    & 0.056    & 0.12     & 0.018    \\
\multicolumn{1}{|c|}{1B vs. 2B}       & -0.026   & 0.042    & 0.09     & 0.007    \\
\multicolumn{1}{|c|}{1C u1 vs. 2C u1} & -0.016   & 0.045    & 0.08     & 0.027    \\
\multicolumn{1}{|c|}{1C u7 vs. 2C u7} & 0.01     & 0.047    & 0.093    & 0.04     \\ \hline
\end{tabular}%
}
\end{table}

In Table \ref{tab:deltas_strat_atif}, the deltas for the TAT and $T_{Offl}$, produced by the Cliff's-delta test for ATIF 0.6 and ATIF 1.2 which compare the strategies are shown. Even though the values in the figure are all considered negligible, as they are below 0.147, there is a clear increase seen in the deltas of the offloading time.

\section{Discussion}

\subsection{Strategy 1}

\begin{figure}[ht]
\centering
\includegraphics[width=1\linewidth]{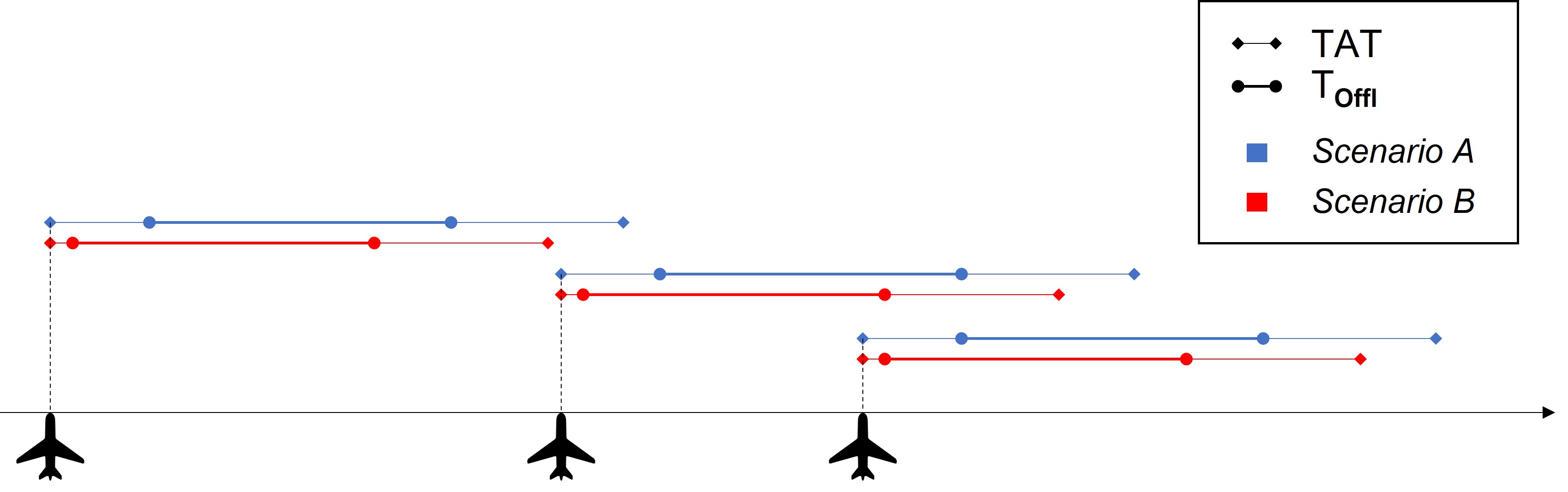}
\caption{Aircraft TAT and offloading time influence.}
\label{fig:ac_schedule_tat_offl}
\end{figure}

For each strategy, in all test cases the offloading time stays more or less the same. In the strategy 1, a factor that could influence a change in offloading time is that the OC does not need to check cargo content. Hence, offloading can start earlier. Comparing the two most extreme cases, scenario A and B, the average time difference for the start of offloading is 18 minutes. This does not mean that at that point in time more resources are available.
This can also be seen in Figure \ref{fig:ac_schedule_tat_offl}, which represents a timeline with three incoming flights. When the offloading of aircraft A is finished, the resources are available for aircraft B. On one hand, in scenario B, the first aircraft has just departed when the second aircraft arrives. The resources are available immediately. 
On the other hand, in scenario A, the first aircraft has not departed while the second aircraft arrives. The offloading of aircraft A has been finished, i.e. that the resources are available to offload the second aircraft. However, the second aircraft still needs pax steps and its content needs to be checked. For the first task, a driver is needed. This takes a longer time as the first aircraft also needs a driver to remove its pax steps. This also contributes to the $T_{BPS}$ increase. 
For the second and third aircraft, it can be seen that even though the TAT is shorter, the time offloading differs. This suggests that the offloading can start earlier and therefore is finished earlier. Nevertheless, this will not make a difference for aircraft B, as it can start offloading earlier as well.

\subsection{Strategy 2}

In strategy 2, the OC can make a resource allocation decision for aircraft A while taking the need for resources for aircraft B into account. This means that even though there are enough resources to start offloading aircraft A, the OC can decide to wait. The OC can also decide to give aircraft A the minimum amount of resources, despite of more resources being available. Yet, these decision are not often made because they are assess by the OC as suboptimal. Furthermore, the OC bases its decision on the earliest finish time for both aircraft A and B, and not the lowest offloading time. 


The decrease in waiting times for increasing certainty in information can be explained by the availability of the OC. When the OC has knowledge of an incoming flight, it has also knowledge of that flight's cargo content. This means that the OC does not need go to the aircraft to check the cargo content. This, on one hand, decreases the waiting time $T_{BC}$. On the other hand, the OC is not leaving its office and is available to react to new tasks thus decreasing the waiting times of other tasks. As other tasks are finished earlier, the resources are free at an earlier point in time. This in its turn, also makes it possible to start a new task sooner.\\

Only three scenarios with unannounced flights are evaluated: one large aircraft, all large aircraft, all aircraft. In the results one unannounced aircraft did not have a visible effect on the TAT. The reason is that the OC only needed to leave its office once, so the effects on other waiting times because of an absent OC is not as strong. Seven unannounced aircraft do have a clear effect on the average TATs and the average waiting times. As the scenarios in between one and seven unannounced aircraft were not investigated, the amount when the effects become significant is not known.


\begin{figure}[ht]
\centering
\includegraphics[width=0.8\linewidth]{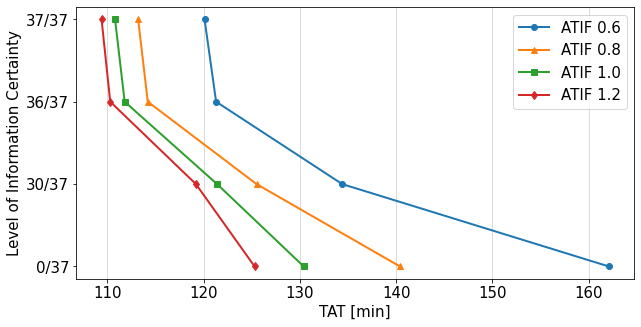}
\caption{TAT trend for the levels of information certainty, for all strategy 1 with all ATIFs. Level of information certainty represents the known number of incoming aircraft over the total number of aircraft. }
\label{fig:tat_atif_strat1}
\end{figure}

\subsection{Arrival time interval factors}

The results of the tests with different ATIFs suggest that the waiting times are more affected than the offloading times. Analysing the results of the different ATIFs, the TAT increases with decreasing interval time. This is shown in Figure \ref{fig:tat_atif_strat1}, where the TAT for all test cases of strategy 1 are shown for all different ATIFs. For all levels of information certainty, the TAT increases considerably the smaller the ATIF becomes. However, the increase in offloading time is substantially less than the increase in TAT. The results of the test cases with different ATIFs indicate that it is mainly the waiting times that increase the TAT. 
The reason is the same as for the main results: as the OC is more often at its office, the OC is able to take quicker action. Since the arrival time interval has been decreased, quicker actions are needed more often. 

The difference in results between the strategies is negligible for all ATIFs. However, the increase in the deltas in Table \ref{tab:deltas_strat_atif} indicate that further increasing the ATIF result in a noticeable difference in offloading time between the two strategies. This can be explained by the airport being too crowded. Waiting for more resources to become available in order to have faster offloading will always take longer than settling for the available, and probably, minimum amount of resources. This is most likely also the reason why the offloading times increase when the airport is more crowded.

It can be concluded that in the case study, the effects of anticipation on the performance of the cargo handling are negligible. However, this might have an effect when the airport becomes less crowded. 
Furthermore, the results suggest that one unannounced aircraft does not have a clear effect. However, the less information is known about incoming aircraft, the higher all the waiting times. Moreover, in this case study under these particular scenarios and circumstances the variation in offloading times can be neglected.

\subsection{Resilience}

This study investigate the resilience of Saint Martin at two levels. The first level is the resilience of the Island after the hurricane disaster. Many buildings and infrastructures were damaged, and the recovery process has been effective due, among other, to ports and airports. Goods, food, and other resources have be carried by ships and planes and delivered to the island. Consequently, efficient airport operations contributed to the island community resilience. The second level is the airport level. Ongoing cargo and passengers handling operations can be significantly disturbed by unannounced incoming flights. These flights need to receive instructions regarding the use of runway, taxiways, parking spots, etc. The airport performance, which corresponds to the turnaround time in this study, can be impacted by unannounced incoming aircraft. The officer coordinator needs to manage these aircraft, and adapt plans to the new information. 

Resilience can be seen as capacity to anticipate undesirable situations and to be ready to efficiently respond to unknown events \cite{su132413915}. These unknown events can create disturbances in the system's organisation. This corresponds to the airports cargo handling operations as analysed in the study: the OC can make a prediction regarding the resources needed by an incoming flight to unload its cargo and handle the passengers that need to be evacuated. The OC needs to adapt the resources allocation schedule in real time to any new unplanned flights. On the Figure \ref{fig:tat_atif_strat1}, the average turnaround times are presented with respect to the number of unknown incoming flights. The offloading operations are only a little impacted by one unannounced aircraft: the average offloading times remain constant. However, when more disruptions occur (7 or 37 unplanned flights), the system's performance decreases. This corresponds to an increase of the turnaround time. The system extends its handling capacity until the decompensation occurs. At that stage, the system has exhausted its adaptive capacity and the performance gracefully degrades. There is no collapse of the system in the investigated scenarios (i.e. the situation where aircraft cannot be unloaded), however an ATIF significantly higher could make the airport performance collapse. This would be the case in the following conflicting situation: when incoming aircraft need to use the runway to land but there is no parking spots available, and at the same time aircraft on the ground need to use the runway to takeoff.

\subsection{Future work}
Although this research aimed to be as close to reality as possible, assumptions needed to be made. One promising avenue for addressing computational challenges and enabling more extensive scenario testing is the use of surrogate modeling techniques within ABMS, as explored in \cite{surrogateABM}. For one, the evacuation operations are modelled as a time penalty. The evacuation operations are likely to cause more delays. Secondly, it is assumed that when the OC has knowledge of the incoming flight, that knowledge is always correct. Unfortunately, in a real situation that is not always the case. The same applies to the communication between all agents. In reality miscommunications happen, especially in such a complex system. Furthermore, in the future, the amount of unannounced flights could be varied to examine which amount of unannounced flights give a significant results. In addition, different case studies could be tested, as well as the effects of miscommunication on TAT of the aircraft.  To study the influence estimation and anticipation have on the performance of the system, the ATIF could be increased even more. Additionally, other anticipation strategies, e.g., optimizing for offloading time, could be examined as well.



\section{Conclusions}
\label{sec:Concl} 



In this research a base model was developed based on van Liere's model. After analysing the case study it was found that the offloading strategy used in van Liere's model was unrealistic. According to experts a realistic strategy is to use as many personnel and GSE as available, to offload an aircraft. In case the next aircraft to offload is known, this aircraft should be taken into account. This is because the main goal of the cargo handling coordinators is to minimize the TAT. In order to evaluate a change in TAT, the TAT can be split in three main parts: the offloading time, the boarding time and the cumulative waiting time.




In this paper, the evaluation of the performance of the TAT was performed by testing different scenarios while implementing the identified strategies. The conclusions for the effects of incomplete knowledge of incoming flights are first of all, the effect of one unannounced aircraft on the total average TAT is negligible. If all seven arrive unannounced the TATs are visibly increased. As the scenarios in between one and all seven are not tested, the exact amount that causes a significant change in the TAT is not known. Secondly, in all test cases the difference in offloading time and boarding time is also negligible, meaning that the waiting times are influenced the most. With decreasing knowledge of incoming flights, the waiting times increase. As the OC needs to walk to the aircraft to check the content, the waiting time of that aircraft increases. At the same time, the OC is not at its office and cannot react to new tasks. 

Additionally, it can be concluded that the difference between the two resource allocation strategies are negligible for this particular case study. When changing the arrival interval time between the aircraft, these effects were still insignificant. However, the deltas of the Cliff's-delta test do increase when increasing the arrival interval time. This suggests that when this arrival interval time is widened more, an effect might be noticeable between the two strategies. Nevertheless, the results suggest that, for this case study, the maximum available resources strategy would have been sufficient.

\section*{Abbreviations}

The following abbreviations are used in this manuscript:

\vspace{1em}
\begin{tabular}{@{}ll}
\textbf{ABM}   & Agent-Based Model \\
\textbf{AC}    & Aircraft \\
\textbf{ATC}   & Air Traffic Control \\
\textbf{ATIF}  & Arrival Time Interval Factor \\
\textbf{GSE}   & Ground Support Equipment \\
\textbf{ICU}   & Individual Cargo Unit \\
\textbf{MOVCON}& Movement Control \\
\textbf{OC}    & Offloading Coordinator \\
\textbf{PJIA}  & Princess Juliana International Airport \\
\textbf{TAT}   & Turn-Around Time \\
\textbf{ULD}   & Unit Load Device \\
\end{tabular}





\bibliographystyle{ACM-Reference-Format-Order}

\end{document}